# Fuzzy Route Switching For Energy Preservation (FEP) in Ad Hoc Networks

**Anuradha Banerjee[1], Paramartha Dutta[2] and Abu Sufian[3]**

[1]Department of Computer Applications, Kalyani Govt. Engg. College, Nadia, Kalyani – 741127, West Bengal, India; anuradha79bn@gmail.com
[2]Department of Computer and Systems Science, Visva-Bharati University, Bolpur – 731235, West Bengal, India; paramartha.dutta@gmail.com
[3]Department of Computer Science, University of Gour Banga, Maida – 732103,West Bengal, India; sufian.asu@gmail.com

## Abstract

Nodes in ad hoc networks have limited battery power. Hence they require energy-efficient technique to improve average network performance. Maintaining energy-efficiency in ad hoc networks is really challenging because highest energy-efficiency is achieved if all the nodes are always switched off and energy-efficiency will be minimum if all the nodes are fully operational i.e. always turned-on. Energy preservation requires redirection of data packets through some other routes having good performance. This improves data packet delivery ratio and number of alive nodes decreasing cost of messages.

## 1. Introduction

A mobile ad hoc network is a collection of mobile nodes that communicate with each other without any centralized infrastructure. It is useful in the situations which demand emergency rescue or when a war breaks out[1-28]. The communication in ad hoc networks is either single hop or multi-hop. If a node nj resides within the radio range or transmission range of another node ni, then nj directly receives messages from ni in single hop. Otherwise, a chain of routers will be required to convey the message of source to the destination. A lot of routing protocols have been developed for ad hoc network nodes. They select an optimal route based on certain criteria, for communication from source to destination. Whenever any of these routes break, the routing protocols inject route-request packets to discover a new route to the destination. This greatly increases message cost in the network which, in turn, greatly consumes the battery power in nodes forcing them to die early.

Route switching at zero cost is the idea proposed in the present article. At the beginning of a communication session, route discovery and best route selection takes place as per the underlying routing protocol. Front End Processor (FEP) advices to send information (i.e. sequence of routers) about the best, second as well as third best route from source to destination, to the source back embedded in route-reply messages. Moreover FEP also introduces fuzzy controlled sleep-request sleep-grant mechanisms that consider various aspects in terms of node behavior and current communication scenarios in the network before granting sleep to a node. Also the sleep duration is not equal for all at all the time. It is computed depending upon those mentioned factors.

## 2. Related Work

A lot of energy aware routing protocols have been proposed in the literature for ad hoc networks. In reference[2], Maximum Residual Packet Capacity (MRPC) routing is proposed. It computes possible number of packets that may be transmitted through each route. The optimal is the one who tops the list that is able to transfer maximum number of packets. Another protocol, Minimum Battery

* Author for correspondence

Cost Routing (MBCR)[3] elects a route with maximum remaining battery capacity. In the Min-Max Battery Cost Routing (MMBCR)[4] battery power of a route is equal to the minimum residual battery power of a node within the route. If there is a tie, MMBCR chooses the route with the shortest hop count. When all nodes in the network have almost identical residual battery power, MMBCR would result in frequent route changes. On the other hand, the minimum transmission power routing MTPR[5] considers the summation of energy consumed per hop as the metric. The route that produces the least sum of transmission power of senders is elected for communication.

Yuan Xue et. al. proposed a location aided power-aware routing protocol in ad hoc networks in[6]. It reduces the transmission power of senders in each hop so that the signal just reaches the receiver. It uses a greedy algorithm to determine the relay region of neighbors and the region in which the desired next hop destination is located, is flooded with that signal. Load prediction routing (LPR[7]) is another state-of-the-art routing protocol which utilizes min-max strategy for energy preservation. However, it ignores the factors like history of communication in a given link, forwarding attitude of individual nodes etc.

Certain energy conservation schemes have been proposed earlier in literature[8-10,21,24-26]]. Some are based on the concept of adjusting radio-range during message transmission depending upon the distance between the sender and receiver in a HOP[8]. Some others select the route with the maximum value of minimum node energy among all the routers[9]. Reference[10] considers the residual energy of nodes, distance from destination and call arrival rate as metrics. However, these are all energy-aware routing protocols. In PEN[22], the nodes operate and communicate in an asynchronous manner. No master node is required and therefore the costly procedure of master selection as well as overloading can be avoided. However, nodes have to periodically wake up, broadcast beacons to say that it is up now and listen eagerly to what the neighbors want to say to it. The protocol STRC (Self Adjusting Transmission Range Control[23]) is based on adjusting of variable transmission range of mobile hosts that is supposed to protect communication links and preserve energy of network nodes used for route discovery as a result of link breakage. But since there exists a practical upper limit of radio-range of any node, link breakages occurs frequently unless the movement of neighbors of a node is taken care of.

Except these energy efficient protocols, there are proactive and reactive routing protocols. Among proactive routing protocols, Destination-Sequenced Distance Vector (DSDV)[12], Cluster-based Gateway Switch Routing (CGSR)[13], Global State Routing (GSR)[14] are well-known. These protocols store route information to virtually every other node in the network. Hence they require a regular update of its routing tables which uses up a significant amount of battery power as well as bandwidth even when the network is idle. Among reactive routing protocols ad hoc on demand distance vector routing (AODV)[15,27,28], Flow Oriented Routing Protocol (FORP) [11], Associativity-Based Routing (ABR)[17] and Fuzzy Controlled Adaptive and Intelligent Route (FAIR)[1] are mentionworthy. Route building is performed on-demand through a route-request, route-reply cycle. Among all the paths through which a route-request packet reaches the destination, the one that is most suitable (according to the performance matrices of the individual protocols), is used for data packet communication.

## 3. FEP in Detail

### 3.1 Overview of FEP

Future Energy Performance (FEP) is an energy-preservation technique in ad hoc network nodes, which is independent of the underlying reactive routing protocol. FEP requires that the source node store multiple routes to each of the destination nodes with whom the source has live communication sessions. Once a node in one such route feels exhausted due to the scarcity of battery charge (present battery charge is less than 40% of total battery charge) or due to excessive packet forwarding load (packet arrival rate is higher than packet departure rate), it sends a sleep-request message to all of its uplink neighbours. Selection of these uplink neighbours is performed by a fuzzy-controller Second Life- Requirements (SL-REQ). Each of those uplink neighbours examine whether sleep can be grant to that node. Those who grant sleep stop forwarding packets to that sleepy node for a predefined sleep time L. Working principle of SL-REQ is based on the following heuristics:

- If packet arrival rate from an uplink neighbor $n_a$ of $n_b$ is higher than the average packet arrival rate at $n_b$, then $n_a$ should consider redirecting its traffic avoiding $n_b$, for a certain period of time. The predefined upper limit of sleep that can be allowed to any downlink neighbor per shot is denoted by L.

- If a downlink neighbor $n_b$ of $n_a$ has transmitted most of the packets sent to it earlier from $n_a$ and has not asked for sleep too many number of times, then $n_a$ should relieve $n_b$ for some predefined time from packet forwarding considering $n_b$ to be asleep although $n_b$ may actually be operating at that time forwarding data packets of some other uplink neighbors.
- If most of the data packets to be transferred through the hop $n_a \rightarrow n_b$ in different communication sessions, have already been transferred then it is evident that $n_b$ has performed its packet forwarding job up to a great extent. So, $n_a$ may allow $n_b$ to go to sleep for some time, provided alternative routes with good performance exist for most of the live communication sessions through the hop $n_a \rightarrow n_b$.

SL-REQ is invoked by at least one of the two enable inputs, namely, residual energy quotient and packet overload quotient. These are termed as enable inputs. Block diagram of SL-REQ appears in Figure 1. In the heuristics mentioned above, the words and phrases like 'most', 'too many', 'great extent', 'good performance' etc have been used significant number of times. This encourages to model the solution using fuzzy logic. Fuzzy logic is flexible, easy to understand and tolerant of imprecise data. It is based on natural language and can efficiently model non-linear functions of arbitrary complexity.

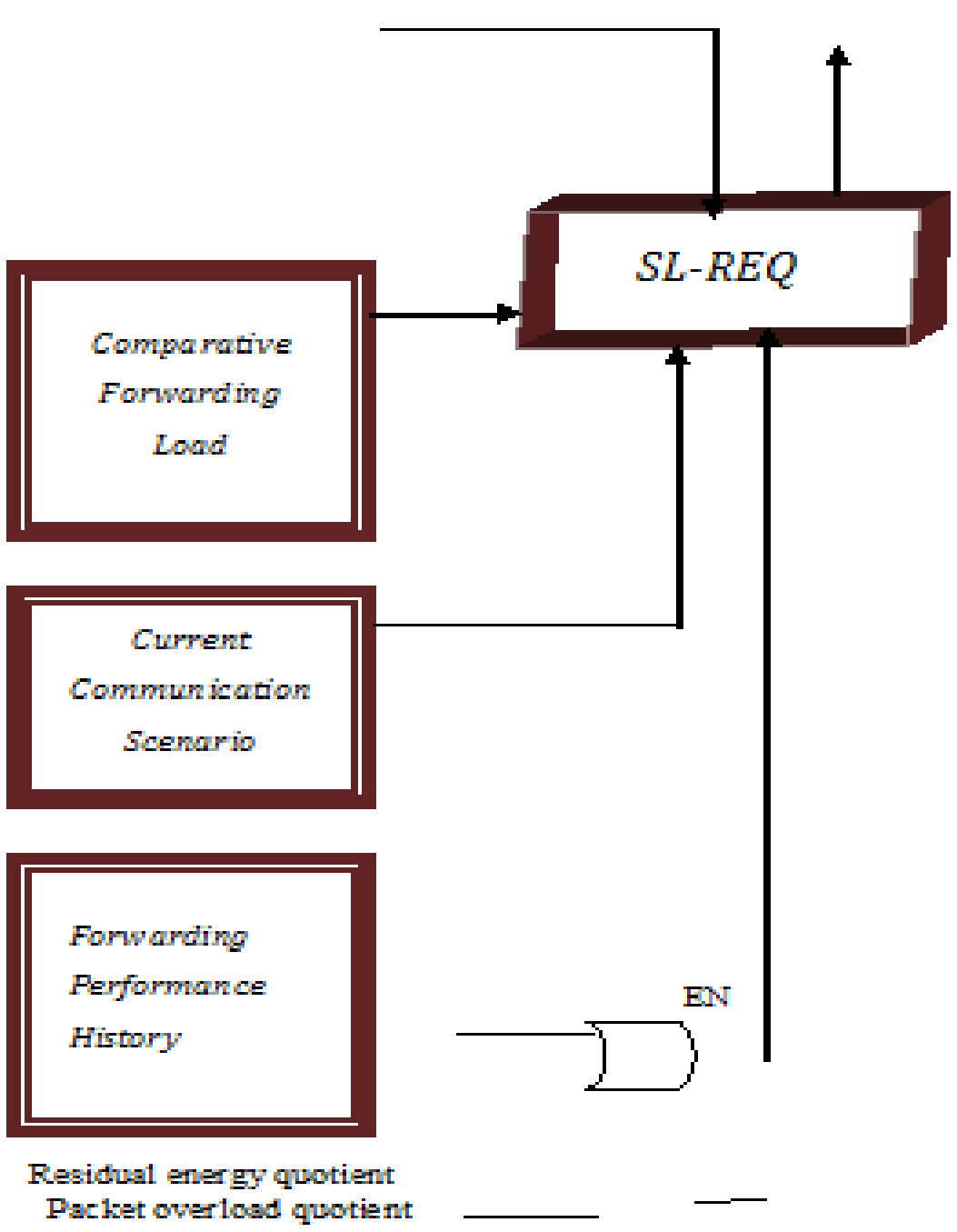

**Figure 1.** Block Diagram of SL-REQ.

## 3.2 Unique Contributions of FEP

Not many routing protocol independent energy saving schemes exist in the literature for ad hoc networks (to the best of author's knowledge). Among them, Performance Excellence Network (PEN) and License Plate Reader (LPR) are noteworthy. But the problem with PEN is huge message cost where each node has to periodically wake up, broadcast beacons and look forward to serve message forwarding requests before really powering down. As already stated in the "related works" section, LPR utilizes the very well known min-max strategy for energy preservation. Keeping in mind all these, the contributions of the present scheme FEP are listed below:

- It is a routing protocol independent energy preservation scheme that uniquely considers various important aspects of node behavior which includes how many packets sent by the current sender have already been forwarded by the current router, how many times sleep has been requested for and granted, performance of alternative routes etc.
- Application of fuzzy logic is very important here since it keeps parity with the behavior of nodes in ad hoc networks. It is discussed in detail in "Overview of FEP" subsection.
- Route switching in FEP does not acquire any extra cost. Whenever a route breaks, the source can try an alternative option without initiating a new route discovery session.

## 3.3 Parameters of SL-REQ

Input parameters of SL-REQ of $n_b$ with respect to one of its uplink neighbor $n_a$ at time t, are as follows:

1. $cl_{a,b}(t)$ – this indicates the comparative message forwarding load at $n_b$ produced by $n_a$ at time t, compared to other uplink neighbors of $n_b$ at that time. It is formulated as,

$$cl_{a,b}(t) = \tau_{a,b}(t) / \tau\text{-avg}_b(t) \qquad (1)$$

where $\tau\text{-avg}_b(t) = \sum_{n_c \in U_b(t)} \tau_{c,b}(t) / |U_b(t)|$

Let $\tau\text{-min}_b(t) = \underset{n \in U_b(t)}{\text{Min}} (\tau_{c,b}(t))$ and $\tau\text{-max}_b(t) = \underset{n_c \in U_b(t)}{\text{Max}} (\tau_{c,b}(t))$

From (1) it can be seen that $cl_{a,b}(t)$ ranges between $(\tau\text{-min}_b(t)/\ \tau\text{-avg}_b(t))$ and $(\tau\text{-max}_b(t)/\ \tau\text{-avg}_b(t))$. If $cl_{a,b}(t)$ is high i.e. it is close to $(\tau\text{-max}_b(t)/\ \tau\text{-avg}_b(t))$ then it indicates that compared to other uplink neighbors of $n_b$ at time t, $n_a$ has transmitted a huge number of packets to $n_b$ for forwarding. Hence, $n_a$ should allow $n_b$ to go to sleep when $n_b$ requires.

2. $ph_{a,b}(t)$ – It is a measure of the packet forwarding performance of $n_b$ with respect to the packet forwarding load produced by $n_a$, as per the history of communication between those two nodes. It is mathematically formulated in (2).

$$ph_{a,b}(t) = (1 - 1/\, sl_{a,b}(t)) \exp (1 - r_{a,b}(t) / s_{a,b}(t)) \quad (2)$$

$r_{a,b}(t)$ is the total number of packets of $n_a$ transmitted by $n_b$ till the current time t. $s_{a,b}(t)$, on the other hand, is the total number of packets sent from $n_a$ to $n_b$ till time t for forwarding. $sl_{a,b}(t)$ is the total number of times $n_b$ has asked for sleep grant from $n_a$.

This parameter of SL-REQ ranges between 0 and 1. If $ph_{a,b}(t)$ is high then it indicates that $n_b$ has successfully transmitted most of the packets sent to it by $n_a$ by the time t. Also $n_b$ has not frequently asked for sleep grant from $n_a$.

3. $ccs_{a,b}(t)$ – It is the current communication scenario of the hop from $n_a$ to $n_b$ at time t. It deals with the number of data packets yet to be transferred in all the live communication sessions passing through that hop.

$$ccs_{a,b}(t) = \sum_{\phi \in R_{a\to b}(t)} (f1(\phi) \exp (1 - f2(\phi)) / |R_{a\to b}(t)| \quad (3)$$

$$f1(\phi) = \{1 / (1 + \alpha 2(\phi) / \alpha 1(\phi))\}$$

$$f2(\phi) = (\sum_{\delta \in \beta 1(\phi)} \beta^{avg}{}_2(\delta)) / |\beta 1(\phi)|$$

$$\beta^{avg}{}_2(\delta) = (\beta^{min}{}_2(\delta) + \beta^{max}{}_2(\delta)) / 2$$

$$\beta^{min}{}_2(\delta) = \min_{\delta \in \beta 1(\phi)} ( \beta 2(\delta)) \text{ and } \beta^{max}{}_2(\delta) = \max_{\delta \in \beta 1(\phi)} ( \beta 2(\delta))$$

The set of alive communication routes from $n_a$ to $n_b$ at time t, is denoted by $R_{a\to b}(t)$. For each of those routes $\phi \in R_{a\to b}(t)$, number of data packets $n_b$ has already transmitted, is $\alpha 1(\phi)$ and number of packets yet to be forwarded is $\alpha 2(\phi)$. Let $\beta 1(\phi)$ denote the set of alternative routes with good performance. For each $\delta \in \beta 1(\phi)$, $\beta 2(\delta)$ indicates the fuzzy performance of the path. Here we have considered two state of the art routing protocols Ad hoc On-Demand Vector (AODV) and FAIR. By default, the route performance in FAIR is fuzzified and its crisp value ranges between 0 and 1. But in AODV, route performance is measured in the form of hop count. In this article, we have remodeled the performance metric as (hop count of the route /maximum allowable hop count in the network). Hence it ranges between 0 and 1 and is fuzzified as per the output parameter SLP of SL-REQ in table 1. Similarly, performance of routes in other routing protocols can also be fuzzified. Let $\beta^{min}{}_2(\delta)$ and $\beta^{max}{}_2(\delta)$ are the minimum and maximum values in the range of $\beta 2(\delta)$ as per the output parameter of SL-REQ in Table 1. So, if $\beta 2(\delta)$ is a2, $\beta^{min}{}_2(\delta)$ and $\beta^{max}{}_2(\delta)$ are 0.25 and 0.50 respectively. Similarly, if $\beta 2(\delta)$ is a3, $\beta^{min}{}_2(\delta)$ and $\beta^{max}{}_2(\delta)$ are 0.5 and 0.75 respectively.

**Table 1.** Range Division of Parameters

| Range division of cl | Range division of ph, ccs and SLPR | Fuzzy variable |
|---|---|---|
| (cl-l) – 0.25(3 cl-l + cl-h) | 0 – 0.25 | A1 |
| 0.25(3 cl-l + cl-h) - 0.5(cl-l + cl-h) | 0.25 – 0.5 | A2 |
| 0.5(cl-l + cl-h) - 0.25(cl-l + 3 cl-h) | 0.5 – 0.75 | A3 |
| 0.25(cl-l + 3cl-h) - cl-h | 0.75 – 1.00 | A4 |

If $ccs_{a,b}(t)$ is high, it indicates that $n_b$ has already transferred a huge number of data packets corresponding to the current communication sessions passing through the hop from $n_a$ to $n_b$ till time t and sufficient alternative routes with good performance are already available surpassing $n_b$. So, if $n_b$ is allowed to sleep it won't harm the network communication much. Most important thing is that it does not initiate a new route discovery process. Hence, message cost in the network does not increase.

***Formulation of enable inputs***

1. $e_b(t)$ – It denotes the residual energy quotient of $n_b$ at time t and it is formulated in (4).

$$e_d(t) = \begin{cases} 1 \text{ if } l_b(t)/E_b \text{ is less than } 0.6 \\ 0 \text{ Otherwise} \end{cases} \quad (4)$$

Here $E1_b(t)$ and $E_b$ denote the consumed energy of $n_b$ at time t and the maximum energy of $n_b$ respectively. $e_b(t)$ is 1 when $n_b$ is almost exhausted (40% of maximum charge is required to remain operable).

2. $ol_b(t)$ – It denotes the packet overload quotient of $n_b$ at time t and it is formulated in (5).

$$ol_b(t) = \begin{cases} 1 & \text{if } TS_b(t)/TR_b(t) \text{ is less than } 1 \\ 0 \\ \text{Otherwise} \end{cases} \quad (5)$$

Here $TS_b(t)$ and $TR_b(t)$ denote the packet inter-service time and the packet inter-arrival time of $n_b$ at time t respectively. If $ol_b(t)$ is 1 when $n_b$ is overloaded.

### 3.4 Rule Bases of SL-REQ

Crisp range division of input parameters of SL-REQ is shown in Table 1, along with the corresponding fuzzy variables. Table 2 combines the effects of pH and CCS producing a temporary output temperature. Both are given equal weight because both are concerned about behavior of the node that has requested for sleep grant from its uplink neighbor(s). The behavior is measured in terms of the percentage of the packet forwarded with respect to the number of packets sent for retransmission. pH deals with the history of communication while CCS deals with the packet forwarding behavior of the sleepy node in the live communication sessions along with the availability of alternative paths surpassing that node. Temperature is combined with cl in Table2 producing the sleep which is the output of SL-REQ.cl compares the load produced by one uplink neighbor with the same generated by other uplink neighbors of the node that has requested for sleep grant. From the point of view of whether a node should be allowed to sleep, temp is much more important than cl. The reason behind is that even if an uplink neighbor generates huge forwarding load, it may not allow a sleepy node to go to sleep provided it could not successfully forward most of the previous packets sent to it by the uplink neighbor and asked for sleep grant too many occasions earlier. Sleep is granted if SLPR is A3 or A4.

**Table 2.** Fuzzy Combination of pH and CCS Generating Temp.

| ph® ccs ⁻ | A1 | A2 | A3 | A4 |
|---|---|---|---|---|
| A1 | A1 | A1 | A1 | A2 |
| A2 | A1 | A1 | A2 | A2 |
| A3 | A1 | A2 | A3 | A3 |
| A4 | A2 | A2 | A3 | A4 |

## 4. Mathematical Analysis

### 4.1 Determination of Sleep Duration

Let L denote the upper limit of sleep duration per request and I is the upper limit of the number of times a node can request its uplink neighbors to go to sleep, in each of its life cycles. Also assume that, $slp_i(t)$ is the number of installments node $n_i$ has requested for sleep till time t. If more than one uplink neighbors of a node are requested for sleep in one shot, then $slp_i(t)$ will be incremented by 1. Then ni will be able to get another session of nap provided the following two conditions satisfy:

a) $(I - slp_i(t)) > 0$

b) SLPR is A3 or A4

If SLPR = A4 the duration of granted nap is L. On the other hand, if SLPR = A3, sleep duration will be (L × mid(A3) / mid(A4)) where mid is a function that accepts a fuzzy variable A1, A2, A3 or A4 and generates the average of its crisp lower and upper limits. So, mid(A1) = 0.125, mid(A2) = 0.375, mid(A3) = 0.675 and mid(A4) = 0.875. So, if SLPR = A3, nap duration is (L × 0.675 / 0.875) i.e. (L × 0.714).

As far as the loss of sleep is concerned, it is ( L - L × 0.714) i.e. (L × 0.286) in a single installment. So, upper limit of loss of sleep in the entire lifetime of a node is ( I × L × 0.286).

### 4.2 Complexity of FEP

In order to determine SLPR for a node, 5 Table accesses are required: Table 1 is accessed 3 times (once for each input) , and one access for each of the tables 2 and 3. Assuming that N be the total number of nodes in the network, (N-1) is the highest number of uplink neighbors of a node. In one shot, a node can request for sleep from at most all of its uplink neighbors. So, in worst case, complexity per shot is 5(N – 1) i.e. O(N). In the best case, N = 2 and the complexity is 5 i.e. O(1). On the other hand, under uniform node density situation, average number of uplink neighbors of a node is $(\psi 1 \times \pi R^2_{avg})$ where $R_{avg} = (R_{min} + R_{max})/2$. $R_{min}$ and $R_{max}$ are the minimum and maximum radio-ranges in the network. ψ is the uniform node density i.e. the number of nodes per unit area. So, the number of table accesses to determine SLPR, is $(5 \times \psi 1 \times \pi R^2_{avg})$. This computation is based on the assumption that minimum and maximum radio-ranges are equally likely in the network.

**Table 3.** Fuzzy Combination of Temp and cl Generating SLPR

| temp® cl - | A1 | A2 | A3 | A4 |
|---|---|---|---|---|
| A1 | A1 | A1 | A2 | A3 |
| A2 | A1 | A1 | A2 | A3 |
| A3 | A1 | A2 | A3 | A4 |
| A4 | A1 | A2 | A4 | A4 |

## 5. Simulation Results

Simulation of the mobile network has been carried out using ns-2 simulator on 800 MHz Pentium IV processor, 40 GB hard disk capacity and Red Hat Linux version 6.2. Graphs appear in figures 2 to 14 showing emphatic improvements in favor of FEP embedded protocols. Number of nodes has been taken as 50, 100, 200, 400 and 700 in five different independent simulation runs. Speed of a node has been chosen as 5 m/s, 10 m/s, 15 m/s, 25 m/s and 30 m/s in all those simulation runs. Transmission range has been varied between 10m and 50m. Used network area is 2000m × 2000m. Used traffic type is constant bit rate. Mobility models used in various simulations are random waypoint, random walk and Gaussian. Performance of the protocols AODV and ABR have been compared with their PEN, LPR and FEP embedded versions, respectively. The simulation matrices are per node energy consumption (total amount of energy consumed / number of nodes), data packet delivery ratio (total no. of data packets delivered × 100/total no. of packets transmitted), per node message overhead (total no. of messages transmitted / total no. of nodes), end-to-end delay per session (total end-to-end delay/total no. of communication sessions), average number of link breakages per session (total no. of link breakages/ total no. of communication sessions), per node per session cost of repairing the broken links (total no. of messages required to repair the broken links / (total no. of communication sessions × total no. of nodes)) and no. of network partitions. Simulation time was 1000 sec for each run. Maximum sleep duration is 50 ms (L) and maximum number of times sleep can be granted is 10 (I).

FEP greatly balances the message forwarding load in the network by allowing approximately exhausted nodes to sleep whereas some (or all) of its uplink neighbors canalize their busy traffic through some other routes. So, network connectivity remains intact and occurrences of link breakage owing to complete exhaustion of nodes reduce up to a great extent. As a result, irrespective of the underlying protocol, FEP significantly reduces the injection of route-request packets for repairing broken links in the network, as shown in Figure 4 and 5. Since the energy consumed by nodes is directly proportional to the cost of messages, FEP embedded versions of the above mentioned protocols consume much smaller energy to accomplish the tasks similar to the other versions of protocols for example PEN, LPR and ordinary version. This increases the node lifetime (Figure 6 and 7) in FEP embedded protocols and reduces packet collision. As a result, data packet delivery ratio in FEP embedded protocols increase. Figures 2 and 3 graphically illustrate this. However it may be noted in Figure 2 and 3 that initially for all the protocols, packet delivery ratio increases as the number of nodes increase. Reason is the betterment in network connectivity produced by the formation of more links, unless a saturation point is reached. After this saturation point, packet delivery ratio starts decreasing owing to huge message contention, collision and increased energy consumption rate. The amount of this reduction is protocol dependent. It is comparatively smaller for a more efficient protocol.

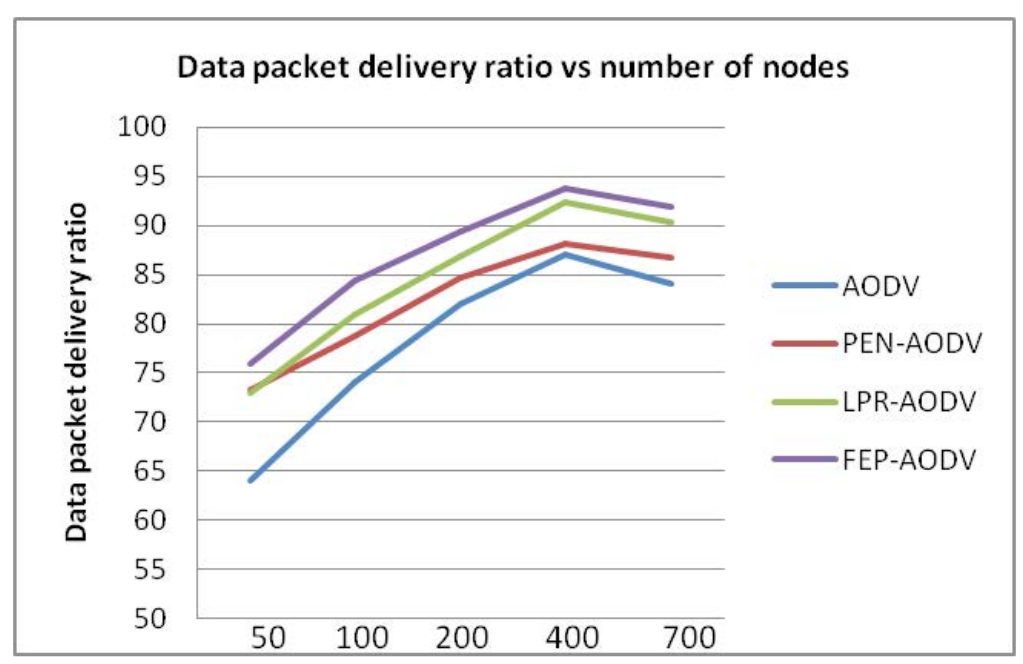

**Figure 2.** Graphical Illustration of Data Packet Delivery Ratio vs. Number of Nodes for AODV.

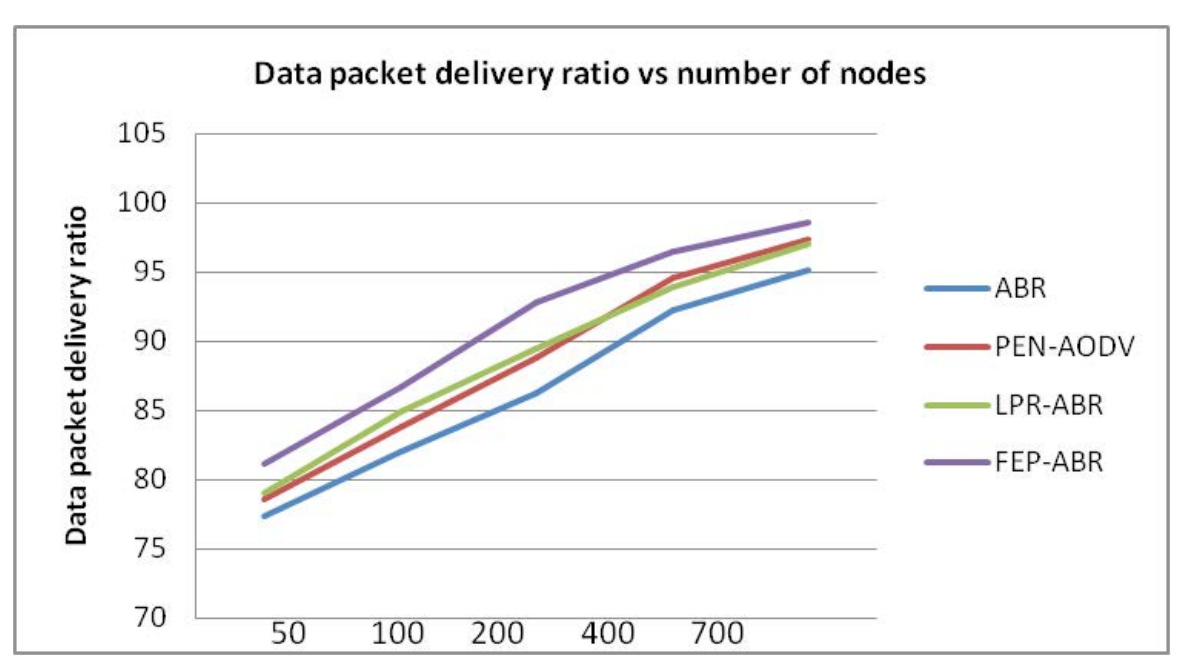

**Figure 3.** Graphical Illustration of Data Packet Delivery Ratio vs. Number of Nodes for ABR.

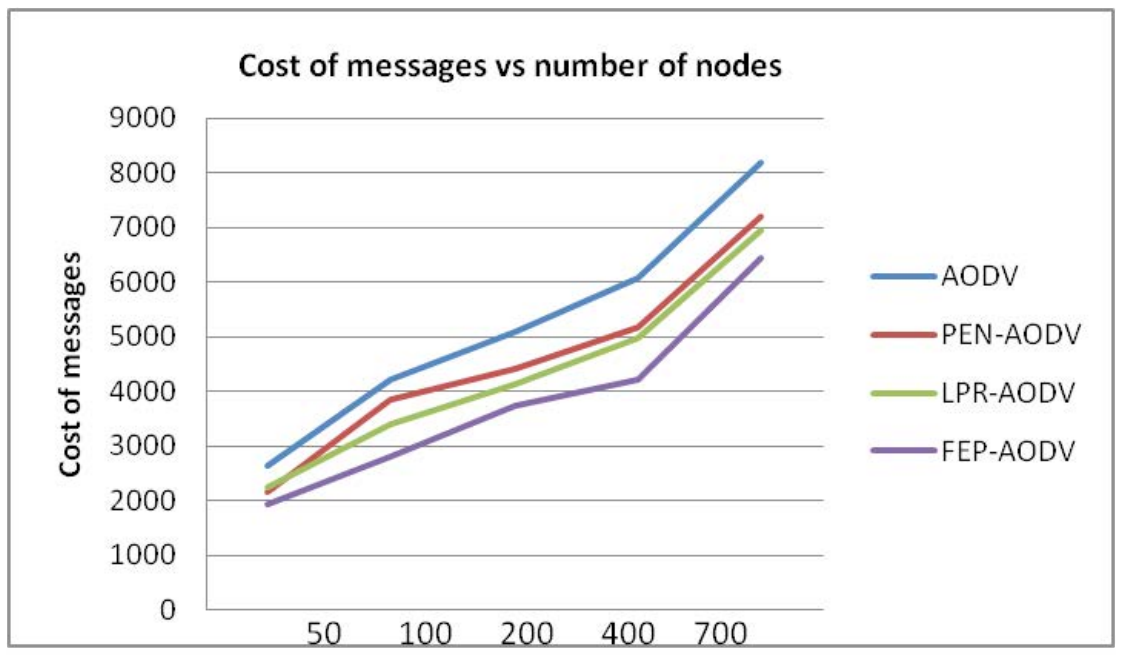

**Figure 4.** Graphical Illustration of Cost of Messages vs. Number of Nodes for AODV.

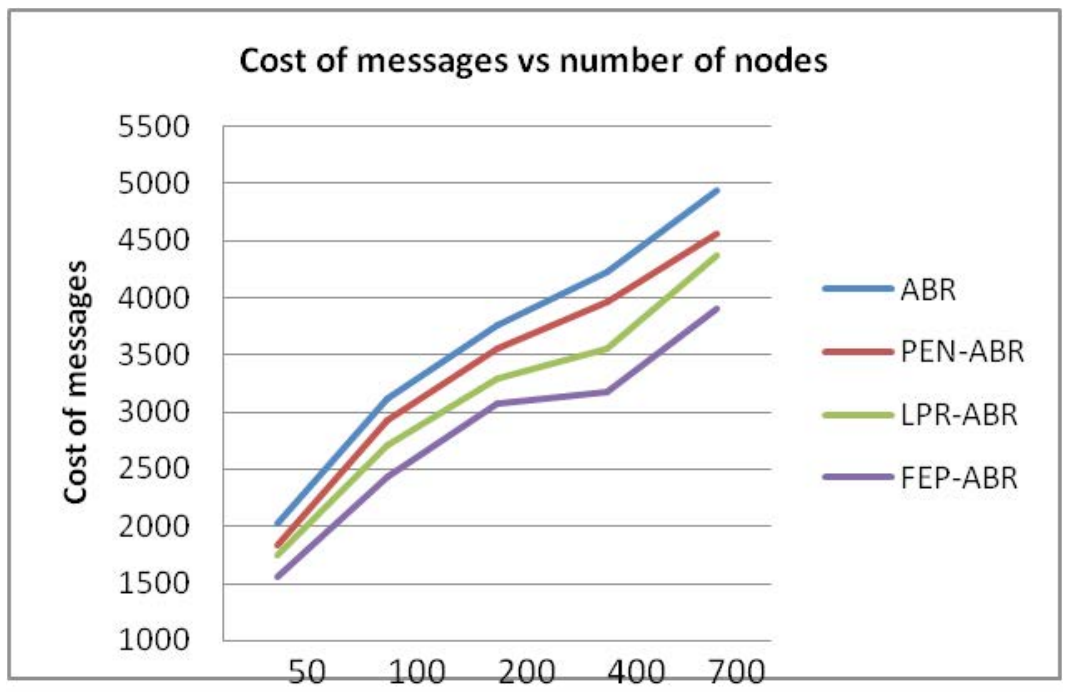

**Figure 5.** Graphical Illustration of Cost of Messages vs. Number of Nodes for ABR.

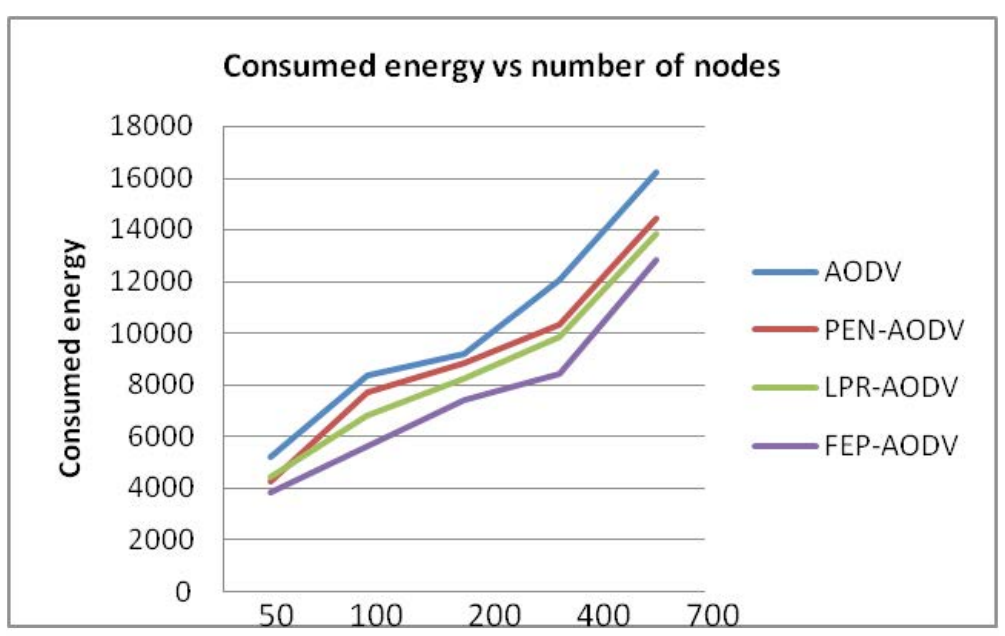

**Figure 6.** Graphical Illustration of Consumed Energy vs. Number of Nodes for AODV.

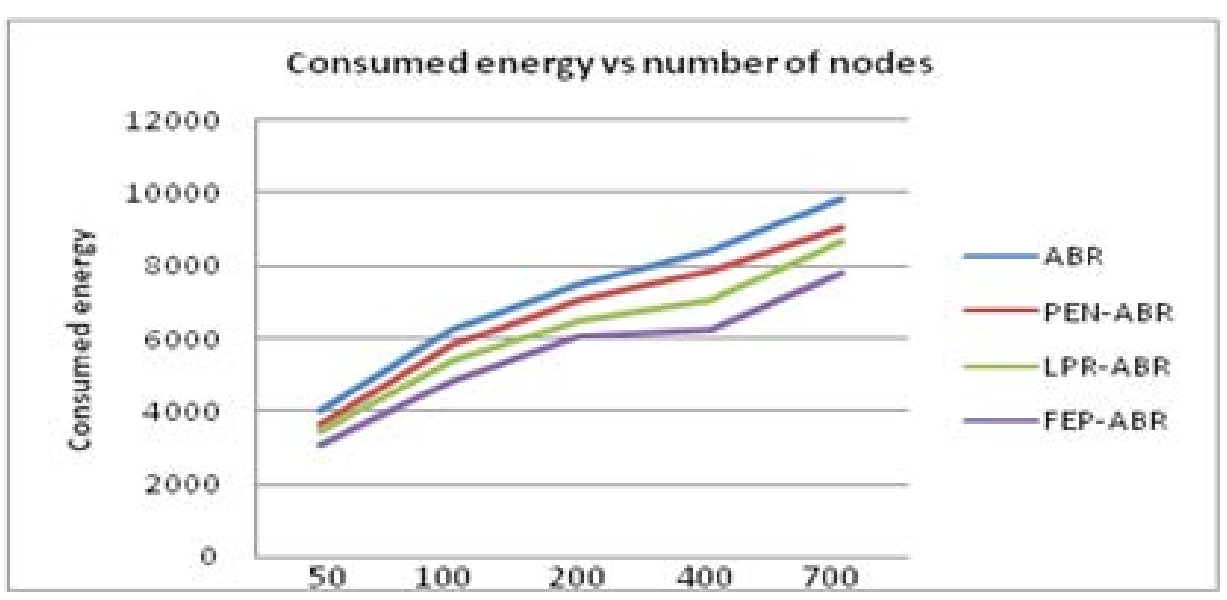

**Figure 7.** Graphical Illustration of Consumed Energy vs. Number of Nodes for ABR.

As far as end-to-end delay is concerned, it increases with message cost. Dead nodes yield link breakages and routes to the destinations of live communication sessions need to be discovered anew. The process of possibly repeated route discovery in the middle of ongoing communication sessions introduce huge amount of delay. As a result, per session end-to-end delay in FEP embedded protocols is much lesser than ordinary protocols. This is seen in Figures 7 and 8.

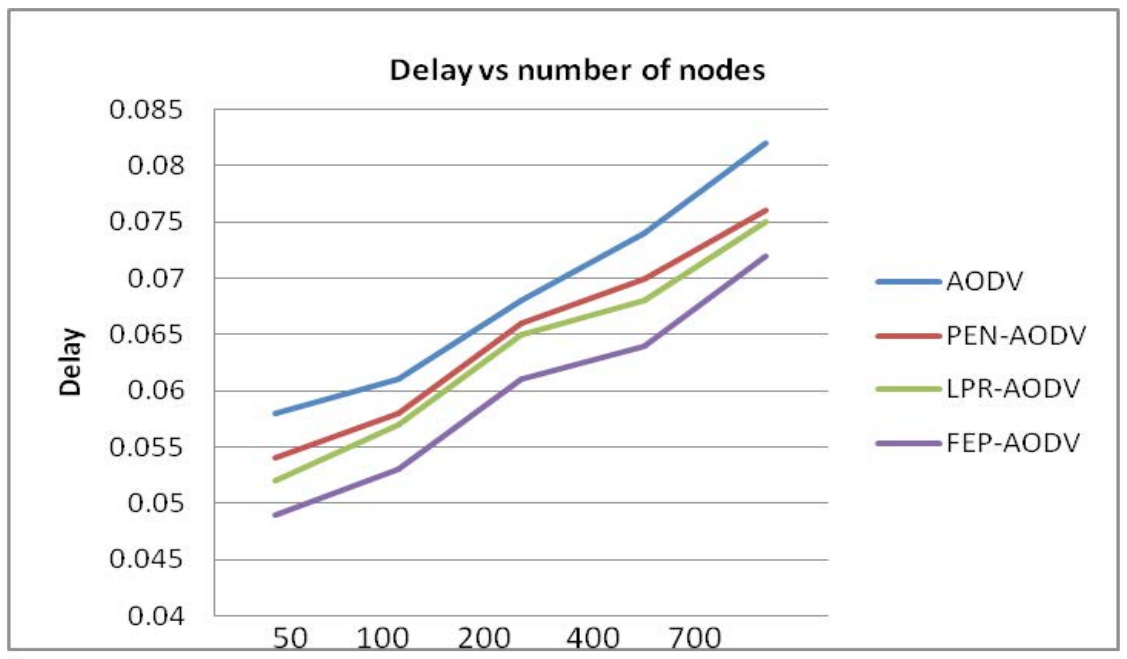

**Figure 8.** Graphical Illustration of Delay Per Session vs. Number of Nodes for AODV.

Figures 9 and 10 graphically demonstrate the average number of links that break within live or ongoing communication sessions. Since FEP allows very exhausted nodes to go to sleep channelizing their forwarding nodes through some other suitable stored alternatives, it prevents most of the link breakages that take place due to the death of routers in an ongoing session. As a result, the average number of link breakages per session is much less in FEP embedded protocols than its competitors. Column type charts have been used for these graphs for better representation; otherwise the lines of various colors would have overlapped in various data points due to the obtained values. Wherever a particular column is absent in a graph, its value is zero. This is applicable for Figures 9 to 14.

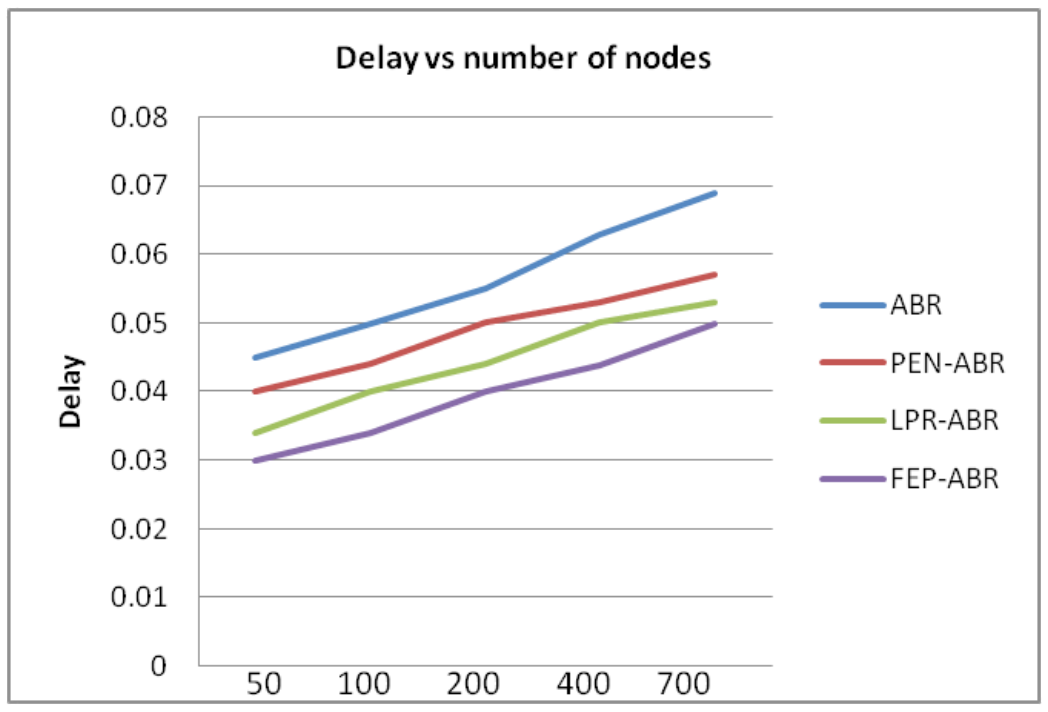

**Figure 9.** Graphical Illustration of Delay Per Session vs Number of Nodes for ABR.

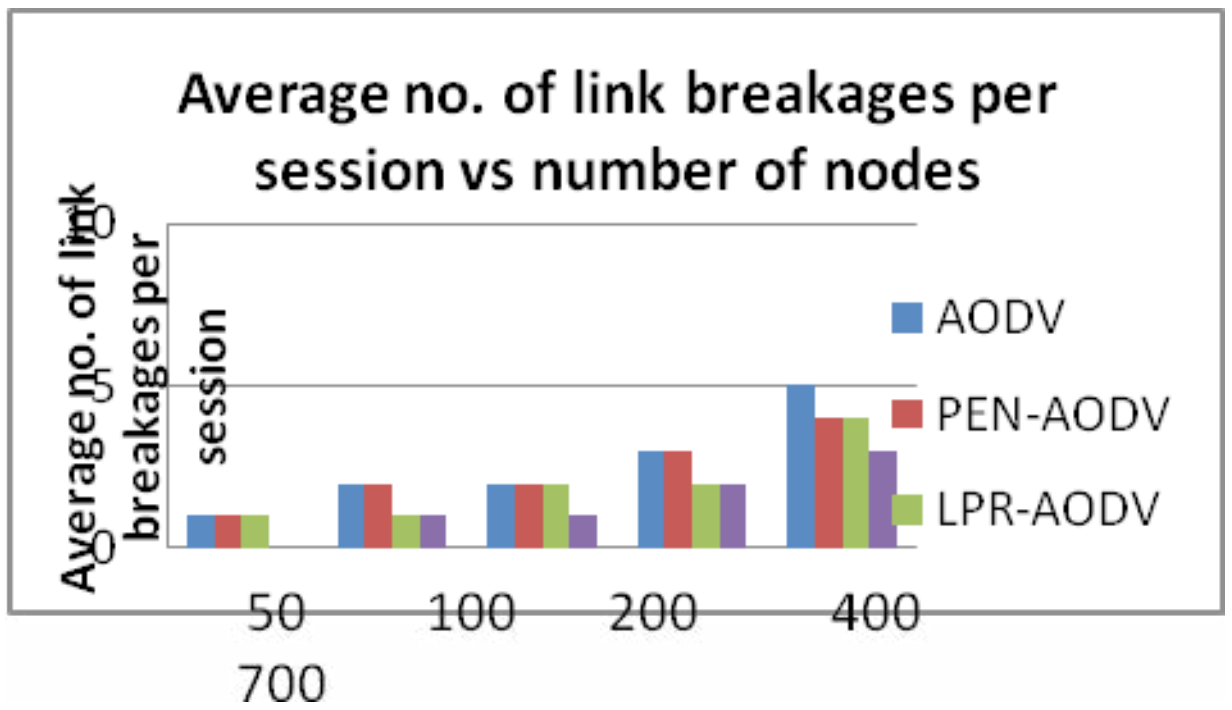

**Figure 10.** Graphical Illustration Of Average No. of Link Breakages Per Session vs. Number of Nodes for AODV.

Figures 11 and 12 deal with the per node per session cost of repairing the broken links. Since FEP proactively stores information about some alternative paths with good performance, the cost of repairing broken routes is much lesser in FEP embedded versions than others. It does not produce any extra cost because this information comes to the source embedded in route-reply packet sent from destinations. Route switching in FEP avoids injecting route-request packets after a link breaks within a session. This saves energy in nodes and in turn, prevents link breakage in other communication sessions within which those nodes play the router's part.

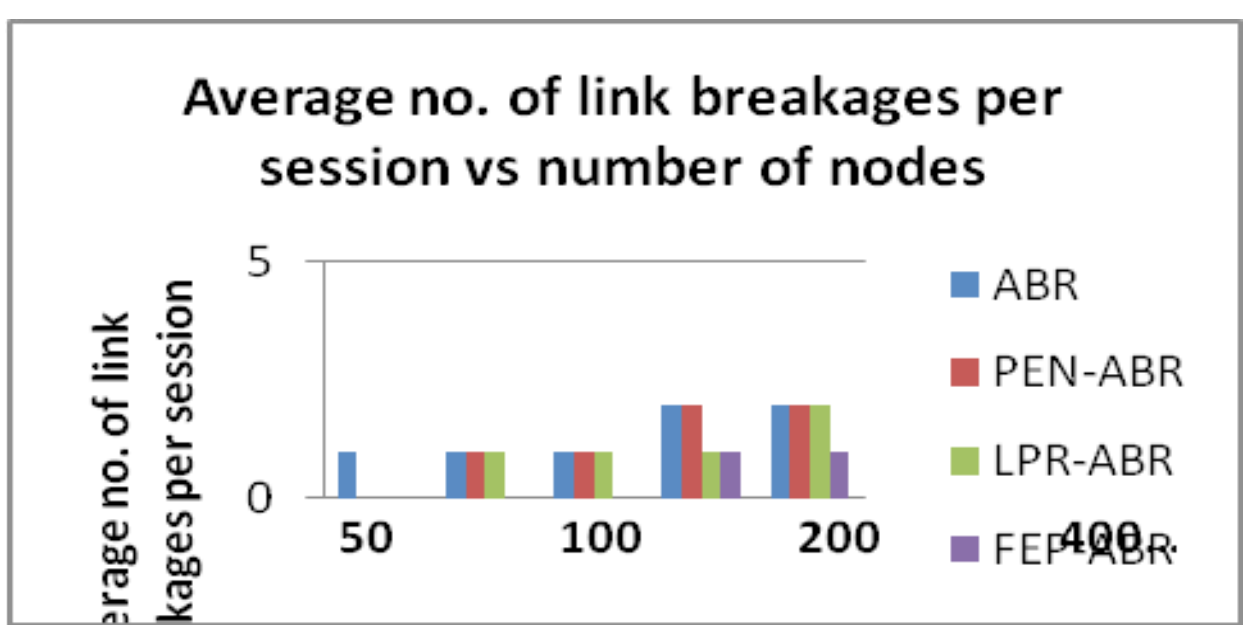

**Figure 11.** Graphical Illustration of Average No. of Link Breakages Per Session vs. Number of Nodes for ABR.

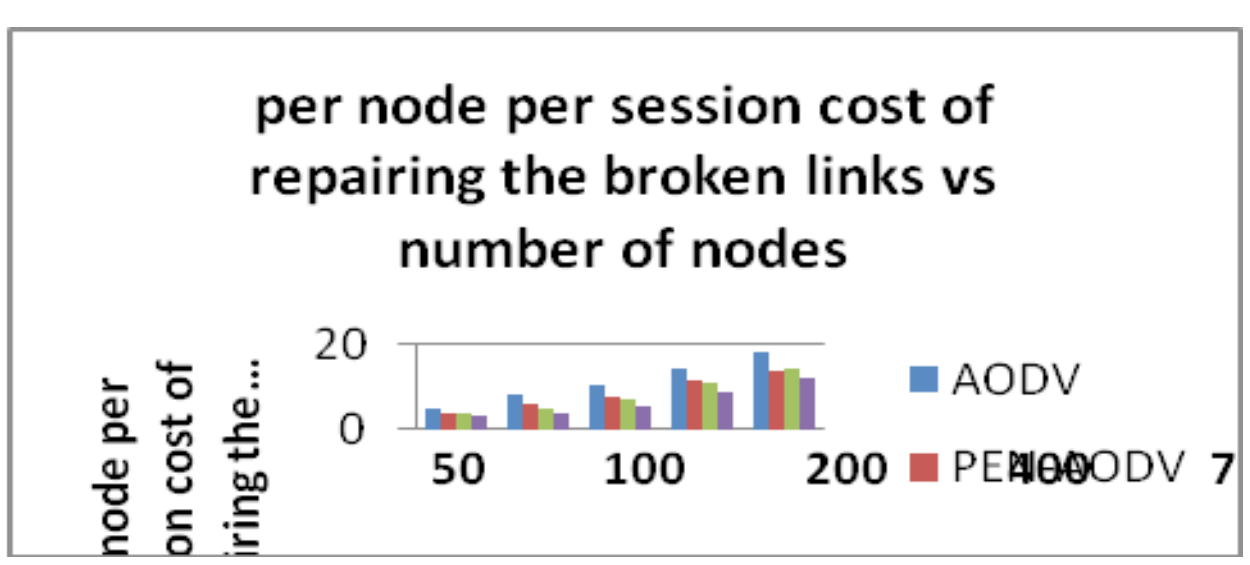

**Figure 12.** Graphical Illustration of Average No. of Per Node Per Session Cost Of Repairing Broken Links vs. Number of Nodes for AODV.

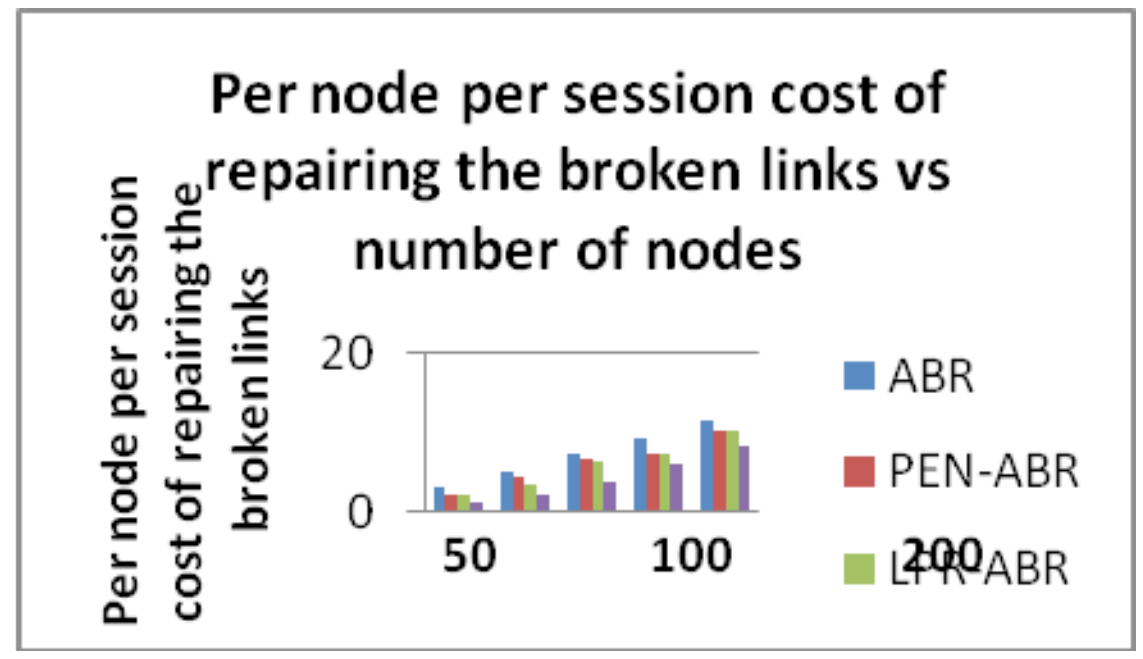

**Figure 13.** Graphical Illustration of Average No. of Per Node Per Session Cost of Repairing Broken Links vs. Number of Nodes for ABR.

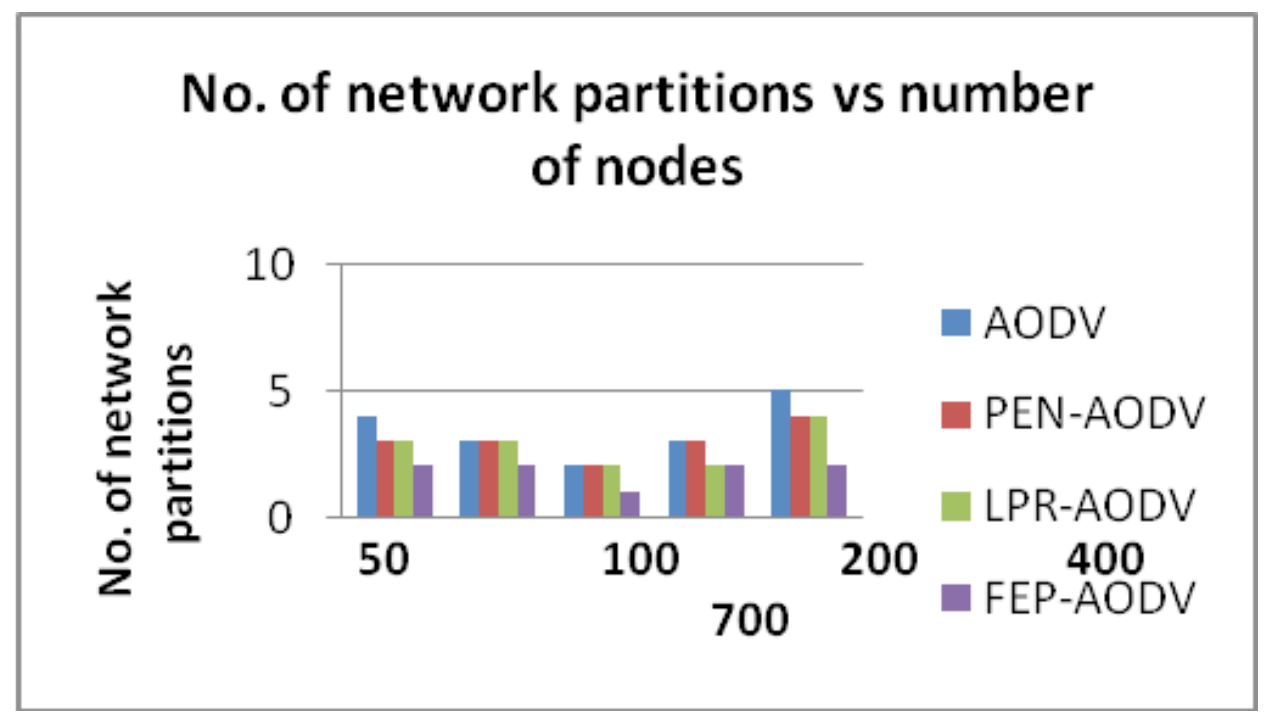

**Figure 14.** Graphical Illustration of Average No. of Number of Network Partitions vs. Number of Nodes for AODV.

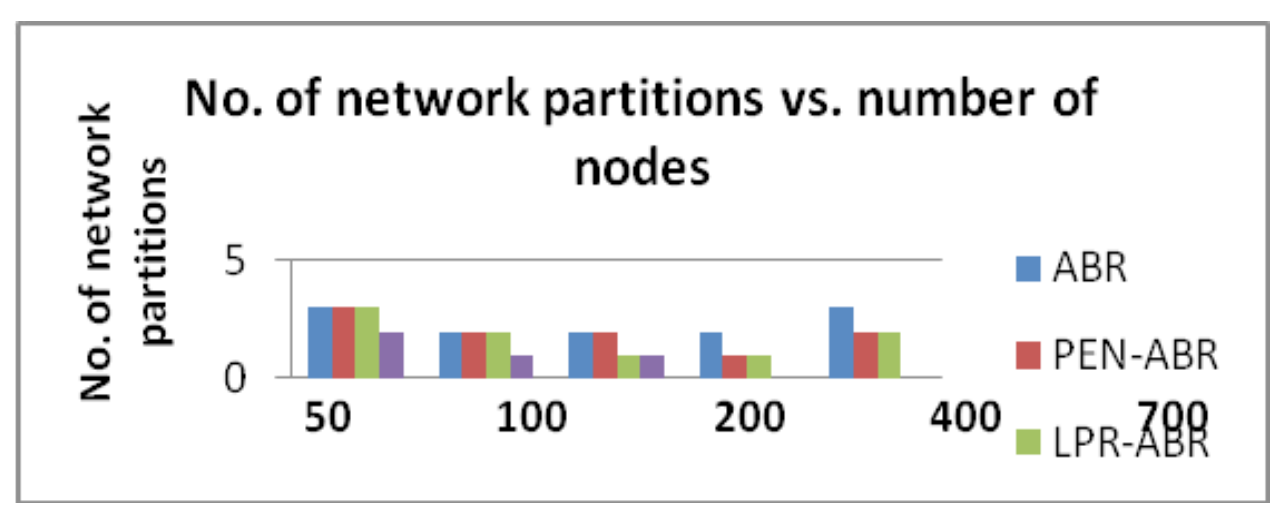

**Figure 15.** Graphical Illustration of Average No. of Number of Network Partitions vs Number of Nodes for ABR.

As far as network partitioning is concerned, when no. of nodes is as low as 50, the network is not well-connected because the nodes are not necessarily closely spaced. This partitioning is not due to node death and energy efficiency doesn't have much to deal with it. So, preventing these partitions is not within the capacity of energy-efficient routing-protocol improvement schemes like PEN, LPR and FEP. As the number of nodes increase to 200 or beyond (400 etc) the improvements produced by FEP becomes more prominent because, at this point, network generally gets partitioned due to excessive power drainage in nodes, not due to lack of connectivity. FEP embedded protocols produce much less number of partitions than others. This is graphically illustrated in Figures 13 and 14.

Improvements produced by FEP embedded protocols over others have been computed in percentage and shown in Table 4. Performance advancement of a protocol X w.r.t. protocol Y corresponding to each number of nodes, have been calculated as [{|X_value – Y_value|)/ MAX(X_value, Y_value)}×100]. In table 4 we can see that

**Table 4.** Improvements Produced by FEP

| Criteria | No. of nodes | FEP-AODV over AODV | FEP-AODV over PEN- AODV | FEP-AODV over LPR-AODV | FEP-ABR over ABR | FEP-ABR over PEN- ABR | FEP-ABR over LPR-ABR |
|---|---|---|---|---|---|---|---|
| Packet delivery ratio | 50 | 21.875 | 6.84 | 6.84 | 10.52 | 6.32 | 6.32 |
| | 100 | 13.33 | 6.25 | 3.65 | 7.5 | 3.61 | 2.4 |
| | 200 | 15.85 | 11.76 | 8.56 | 9.41 | 5.68 | 3.33 |
| | 400 | 12.79 | 7.79 | 4.30 | 9.95 | 6.56 | 6.78 |
| | 700 | 16.16 | 10.22 | 5.31 | 11.1 | 8.45 | 7.93 |
| Per node message cost | 50 | 27.09 | 25.2 | 23.36 | 27 | 24.76 | 24.3 |
| | 100 | 30.23 | 23.05 | 21.285 | 24.67 | 23.79 | 21.71 |
| | 200 | 29 | 19.65 | 21.62 | 23.51 | 23.32 | 23.89 |
| | 400 | 29.02 | 24.67 | 23.24 | 25.53 | 21.56 | 22.18 |
| | 700 | 29.87 | 27.86 | 26.26 | 27.36 | 24.9 | 23.2 |
| Per node energy con-sumption | 50 | 28.75 | 26.41 | 24.4 | 27.4 | 25.2 | 25.1 |
| | 100 | 30 | 25.2 | 23.7 | 24.67 | 24.1 | 21.78 |
| | 200 | 27.98 | 24.07 | 22.91 | 23.5 | 23.33 | 23.4 |
| | 400 | 28.92 | 23.69 | 23.4 | 25.79 | 22.7 | 23.5 |
| | 700 | 30.2 | 28.1 | 27.01 | 28.09 | 25.6 | 24.78 |
| End-to-end delay per session | 50 | 17.24 | 11.11 | 10.09 | 18.67 | 13.15 | 10.13 |
| | 100 | 15.87 | 8.62 | 6.45 | 16.85 | 11.1 | 9.43 |
| | 200 | 13.24 | 10.7 | 8.1 | 13.2 | 10 | 9.1 |
| | 400 | 12 | 11.14 | 6.78 | 13.32 | 10.21 | 10.29 |
| | 700 | 15.66 | 8.87 | 7.4 | 14.1 | 12.5 | 11.56 |
| Average no. of link breakages per session | 50 | 100 | 100 | 100 | 100 | 0 | 0 |
| | 100 | 50 | 50 | 0 | 100 | 100 | 100 |
| | 200 | 50 | 50 | 50 | 100 | 100 | 100 |
| | 400 | 33.33 | 33.33 | 0 | 50 | 50 | 0 |
| | 700 | 40 | 25 | 25 | 50 | 50 | 50 |
| Per node per session cost of repairing broken links | 50 | 38.42 | 25.62 | 25.63 | 67.74 | 50 | 50 |
| | 100 | 50 | 33.33 | 20 | 61.54 | 51.73 | 37.5 |
| | 200 | 44.2 | 24.32 | 22.9 | 48.57 | 43.75 | 41.1 |
| | 400 | 37.76 | 21.23 | 19.09 | 36.26 | 17.14 | 17.3 |
| | 700 | 28.33 | 16.5 | 15 | 28.7 | 18.81 | 18.7 |
| No. of network partitions | 50 | 50 | 33.3 | 33.3 | 33.3 | 33.3 | 33.3 |
| | 100 | 33.3 | 33.3 | 33.3 | 50 | 50 | 50 |
| | 200 | 50 | 50 | 50 | 50 | 50 | 0 |
| | 400 | 33.3 | 33.3 | 0 | 100 | 100 | 100 |
| | 700 | 60 | 50 | 50 | 100 | 100 | 100 |

per node, per session etc. phrases appear over and over again. This emphasizes the improvements demonstrated in Table 4. For eg., when no. of nodes is 700, per node energy consumption reduces by 24.78% in FEP-ABR than ABR. So, if FEP can save almost 1/4 units of energy per node in each simulation run, then overall energy saved in the network is at least (175×e_min) where e_min is the minimum initial battery power of any node considering all nodes in the network. This is extremely significant. Also remarkable improvements have been noticed as far as the metrics "per node per session cost of repairing the broken links", "average number of link breakages per session", and "number of network partitions are concerned".

## 6. Conclusion

This paper proposes a protocol independent fuzzy controlled energy preservation scheme (FEP) that aims at increasing network lifetime and maintaining network connectivity in a busy environment, by intelligently channeling the message forwarding load of busy nodes through nearby idle alternatives. Whenever a node needs to sleep, it asks for that from certain uplink neighbors that produce more load to that node. FEP shows substantial improvement in a busy network environment in terms of packet delivery ratio, agility and reduction in message overhead as well as energy consumption.